\documentclass[12pt]{article}
\usepackage{graphicx}
\usepackage{epsfig}
\begin{document}
\title{From collective rhythm to adaptive synchronization of neural activity}
\author{{
Debin Huang \footnote{E-mail: dbhuang@staff.shu.edu.cn}
}\\\\
{\small Department of Mathematics, Shanghai University, Shanghai
200436, P.R. China}}
\date{}
\maketitle
\begin{center}
\begin{minipage}{12cm}
\hskip 1cm A novel viewpoint, i.e., adaptive synchronization, is
proposed to explore collective rhythm observed in many complex,
self-organizing systems. We show that a simple adaptive coupling
is able to tip arrays of oscillators towards collective
synchronization. Two arrays of simple electrically coupled
Hindmarsh-Rose chaotic neurons are used to illustrate cooperative
dynamics of neural activity like the central pattern generators,
which supplies a new idea for biological experiments and numerical
simulations. The results indicate that such small-world adaptive
coupling may be a universal essence of the collective
rhythm observed in the natural world.\\
\vskip 0.01cm PACS number(s): 87.10.+e; 05.45.-a; 84.35.+i
\end{minipage}
\end{center}
Today one of the main unsolved problems in science is how to
apperceive and study complex, self-organizing systems. A puzzling
characteristic in these systems is the spontaneous collective
rhythm, i.e., so-called collective synchronization. This
remarkable phenomenon has been observed extensively in the natural
world, ranging from inorganic systems to organic systems, e.g.,
Christiaan Huygens' two synchronization clocks, wobbly bridges,
the oscillating uniformly Josephson junctions, the Bose-Einstein
condensate (BEC), emerging coherence in chemical oscillators,
fireflies synchronizing spontaneously their flashes,
self-synchronization of the cardiac pacemaker cells, the
cooperative pattern in network of neurons, animals' gaits, groups
of women with the mutually synchronized menstrual cycles, an
audience clapping in sync, and coherent moving states recently
observed in highway traffic, etc$^{[1]}$. In the collective
rhythm, each individual itself executes two different behaviors,
i.e., periodic oscillating and aperiodic (i.e., chaotic)
oscillating. The collective rhythm of periodic oscillators emerges
in many fields, while the collective chaotic synchronization is
observed in the network of neurons and two weakly coupled BECs,
etc. As is well known, there exist infinitely many unstable
periodic orbits embedded in a chaotic attractor, meanwhile due to
the extreme sensitivity of initial values chaotic systems
intrinsically defy synchronization. Therefore from this sense the
collective synchronization of chaotic oscillators seems to be more
difficult and significant than that of periodic oscillators.
\par
Although collective synchronization has been observed in various
concrete experiments, some key theoretical problems remain open
for scientists, especially for mathematicians. Polymath Norbert
Wiener ever tried to develop a mathematical model of collection of
oscillators in the late 1950s but little fruit was
obtained$^{[2]}$. The theoretical breakthrough came from A.T.
Winfree's pioneering work in 1966$^{[3]}$, where he developed a
mathematical model to study large populations of periodic
oscillators, and obtained some important results although stymied
by the difficulty of solving the mathematical model. A crucial
breakthrough came in 1975 when Y. Curamoto refined and simplified
the mathematical framework developed by Winfree$^{[4]}$, where an
exactly solvable mean-field model of coupled oscillators was
proposed. Since these pioneering studies lots of theoretical works
have been accomplished $^{[5]}$. Note that only the case of
coupled periodic oscillators, i.e., the collective phase
synchronization, was investigated in the literature. Through the
large amount of effort in idealized mathematical models there has
been a common acknowledgement for the complex problem on
collective rhythm: individual oscillators are coupled together,
and there exists a critical value for the intensity of mutual
coupling, below which anarchy prevails and above which coherence
with the collective rhythm reigns. However, in this viewpoint two
points deserve to investigate further. One is how the individual
oscillators are coupled together, and the other is how complex,
self-organizing systems produce the threshold of coupling strength
to reach collective synchronization. In the previous theoretical
work, the weak couplings were almost global, e.g., mean-field
coupling, where each oscillator interacts with the rest of
oscillators. However from the perspective of biology (especially
neurobiology) such all-to-all coupling is too complicated to
benefit to realize promptly the corresponding function. As for the
problem on the critical intensity of coupling, it is still far
away from scientists.
\par
In my opinion, the architecture of interaction of weakly coupling
in the collection of oscillators should be as simple as possible,
and the critical coupling strength should be reached
self-adaptively due to the spontaneousness of collective rhythm.
This speculation is inspired by the provocative  words
``self-adaption creates complexity" which is a basic idea of
complex adaptive systems (CAS) theory, meanwhile this idea is
motivated directly by considering the collective chaos
synchronization. In network of chaotic oscillators, which emerges
widely in the field of neurobiology,  the mathematical model must
be exactly unsolvable, but as is introduced above almost all the
previously obtained results on collective synchronization of
coupled oscillators are based on the solvability of the
mathematical model. In this paper, a simple mathematical framework
is proposed to confirm this speculation, where a simple adaptive
coupling is used to produce the collective synchronization of
arrays of oscillators.\par
Our analysis will be limited to an
array of oscillators that are all strictly identical, which
ignores the diversity. Suppose that the dynamical behavior of each
oscillator is governed by an $n$-dimensional differential equation
$$\dot x=f(x), x\in \texttt{R}^n, \eqno(1)$$
where $f(x)$ is a differentiable nonlinear vector function. We
firstly consider cycle-type coupling in a ring of $N$ oscillators
(note that only the case of unidirectional coupling is considered
here, but the extension of bidirectional coupling is very
natural),
$$\dot x_i=f(x_i)-\epsilon_{i,i-1}(x_i-x_{i-1}), \ \ i=1,2,\cdots, N, \eqno (2)$$
where $x_i=(x_{i,1},x_{i,2},\cdots,x_{i,n})\in \texttt{R}^n$,
$x_0\equiv x_N$,
$\epsilon_{i,i-1}=(\epsilon_{i,i-1,1},\epsilon_{i,i-1,2},\cdots,\epsilon_{i,i-1,n})$,
and
$\epsilon_{i,i-1}(x_i-x_{i-1})\equiv(\epsilon_{i,i-1,1}(x_{i,1}-x_{i-1,1}),
\epsilon_{i,i-1,2}(x_{i,2}-x_{i-1,2}),\cdots,
\epsilon_{i,i-1,n}(x_{i,n}-x_{i-1,n}))$. This model indicates that
the connection topology is cycle-type, and each individual
oscillator is coupled only to its nearest neighbor. Here the
coupling intensity $\epsilon_{i,i-1}$ varies adaptively according
to the following update law (this is very crucial)
$$\dot \epsilon_{i,i-1,j}=\gamma_{i,i-1,j}(x_{i,j}-x_{i-1,j})^2, \ \ i=1,2,\cdots,N,\ \ j=1,2,\cdots,n,\eqno (3)$$
where $\gamma_{i,i-1,j}$ are arbitrary positive constants. Next we
consider another simple configuration of network, i.e., star-type
coupling,
$$\dot x_1=f(x_1),\ \ \ \dot x_i=f(x_i)-\epsilon_{i,1}(x_i-x_1), \ \ i=2,3,\cdots, N, \eqno (4)$$
where all notations are as those above, and the coupling strength
$\epsilon_{i,1}$ varies according to the update law
$$\dot \epsilon_{i,1,j}=\gamma_{i,1,j}(x_{i,j}-x_{1,j})^2, \ \ i=2,\cdots,N,\ \ j=1,2,\cdots,n.\eqno (5)$$
Such star-type coupling admits the characteristics of small-world
networks, where the first oscillator, $x_1$, represents a hub.
\par Using the well-known Lasalle's invariance principle in
mathematics, one may prove that for arbitrary bounded solutions of
system (2)-(3) (or (4)-(5)) as $t\rightarrow\infty$
$x_i(t)\rightarrow x_1(t),i=2,\cdots,N$, and the coupling strength
will converges to a constant dependent on the initial values. The
idea of proof is similar to that in [6]. The results indicate that
the above two coupling schemes can produce the collective
synchronization of $N$ oscillators, and such collective
synchronization is nonlinear stable and robust against the effect
of noise. Moreover the variable coupling strength reaches
self-adaptively a value which just corresponds to the threshold
for arising collective rhythm, meanwhile the small weak coupling
may be obtained by adjusting the dissipation parameter
$\gamma_{i,i-1,j}$. In addition, note that in the present coupling
scheme the mutual interaction will vanish once the collective
synchronization is achieved, which implies that the collective
rhythm doesn't change dynamical behavior of each individual. In
particular, when the oscillator (1) is chaotic, thanks to the
global attraction and nonhyperbolicity of chaotic attractor these
coupling schemes will be more simple and the global collective
synchronization will be realized more easily, see the following
examples on cooperative dynamics in neurobiology.
\par
It has been observed in neurobiological experiments and in
numerical simulations that individual neurons show chaotic
spiking-bursting, while ensemble of such irregularly bursting
neurons can produce cooperatively coherent, rhythmical bursting,
i.e., synchronized pattern of neural activity$^{[7]}$. Such
cooperative property plays a crucial role in neural activity, for
example the central pattern generators (CPG) controlling the
rhythmic motor behavior of animals. An important question is how
neural assemblies produce and control regular rhythm. In the
literature, some global coupling schemes like mean-field coupling
have been used to explore the interesting sync pattern$^{[8]}$.
Here we will attempt to apply the proposed adaptive
synchronization to reveal this complex, self-organizing
phenomenon. To do it, we choose the famous Hindmarsh-Rose neuron
model$^{[9]}$, a third-order ordinary differential equation with
one slow variable modelling chaotic spiking-bursting of neuron, as
dynamical equations of individual neurons. Firstly we consider a
network of neurons as the following cycle-type electrically
coupling scheme,
$$\begin{array}{l}\dot{X}_i=Y_i+3X_i^2-X_i^3-Z_i+I-\epsilon_{i,i-1}(X_i-X_{i-1}), \\
\dot{Y}_i=1-5X_i^2-Y_i,\ \
 \dot{Z}_i=-rZ_i+4r(X_i+1.6),\ \ i=1,2,\cdots, N,
 \end{array} \eqno(6)$$
 with $$\dot\epsilon_{i,i-1}=\gamma_{i,i-1}(X_i-X_{i-1})^2,\eqno (7)$$
where $X_0\equiv X_N$, $r=0.0012$, and the external input
$I=3.281$. Each neuron
 is characterized by three time-dependent variables: the membrane potential $X_i$,
 the recovery variable $Y_i$ and the slow adaptation current
$Z_i$. $\epsilon_{i,i-1}$ represents the self-adaptive strength of
the electric coupling between the $i$th neuron and the $(i-1)$th
neuron. Similarly, the star-type coupling network of neurons with
the first neuron being the hub is given as
$$\begin{array}{l}\dot{X}_i=Y_i+3X_i^2-X_i^3-Z_i+I-\epsilon_{i,1}(X_i-X_1), \\
\dot{Y}_i=1-5X_i^2-Y_i,\ \
 \dot{Z}_i=-rZ_i+4r(X_i+1.6),\ \ i=1,2,\cdots, N,
 \end{array} \eqno(8)$$
 with $\epsilon_{11}\equiv 0$, and otherwise
 $$\dot\epsilon_{i,1}=\gamma_{i,1}(X_i-X_1)^2, \ \ i=2,3,\cdots, N. \eqno (9)$$
\par
For the sake of simplicity, we set the parameters
$\gamma_{ij}\equiv 0.1$ and $N=5$. The adaptive collective
synchronization of the two networks of chaotic neurons is shown
numerically in Figs. 1-4. Comparing the above two coupling schemes
(see Fig. 4 legends), we find that the adaptive synchronization in
the star-type network with the characteristics of small-world is
achieved more easily, which confirms again the strong
synchronizability of small-world networks$^{[10]}$. In conclusion,
such adaptive collective synchronization is more simple than those
based on the global coupling schemes in literature, which supplies
a new idea for biological experiments and numerical simulations.
Moreover, the result supports the well-known neural theory of
high-level brain function proposed by D. Hebb in 1949: Learning
takes place by the adaptive adjustment of the conductances of the
connections between neurons$^{[11]}$.
\par
As the above statements, besides the self-adaption of individual
neurons a key reason why such simple synaptic coupling schemes
(i.e., only coupling one variable representing the membrane
potential) can produce so interesting collective synchronization
of neural activity is the chaotic characteristic of neural
dynamics. This result throws highly light to the well-known
viewpoint that chaos is a necessary ingredient in life. Obviously
the present adaptive collective synchronization can be generalized
to the other networks, such as the multi-star-type networks and
the higher-dimensional lattice networks. In particular, such
adaptive coupling can be extended to the complex networks like
small-world networks and scale-free networks $^{[10,12]}$.
Combining the characteristics of small-world networks, such as
quick signal-propagation and strong synchronizability, etc., we
speculate that such small-world adaptive coupling may be a
universal essence of collective rhythm observed in the natural
world. Therefore we hope that this work will inspire further
studies of complex, self-organizing systems.\par
 {\bf Acknowledgments}
This work is supported by the National Natural Science Foundation
of China (10201020; 10432010).

\newpage

\begin{figure}
\begin{center}\epsfig{file=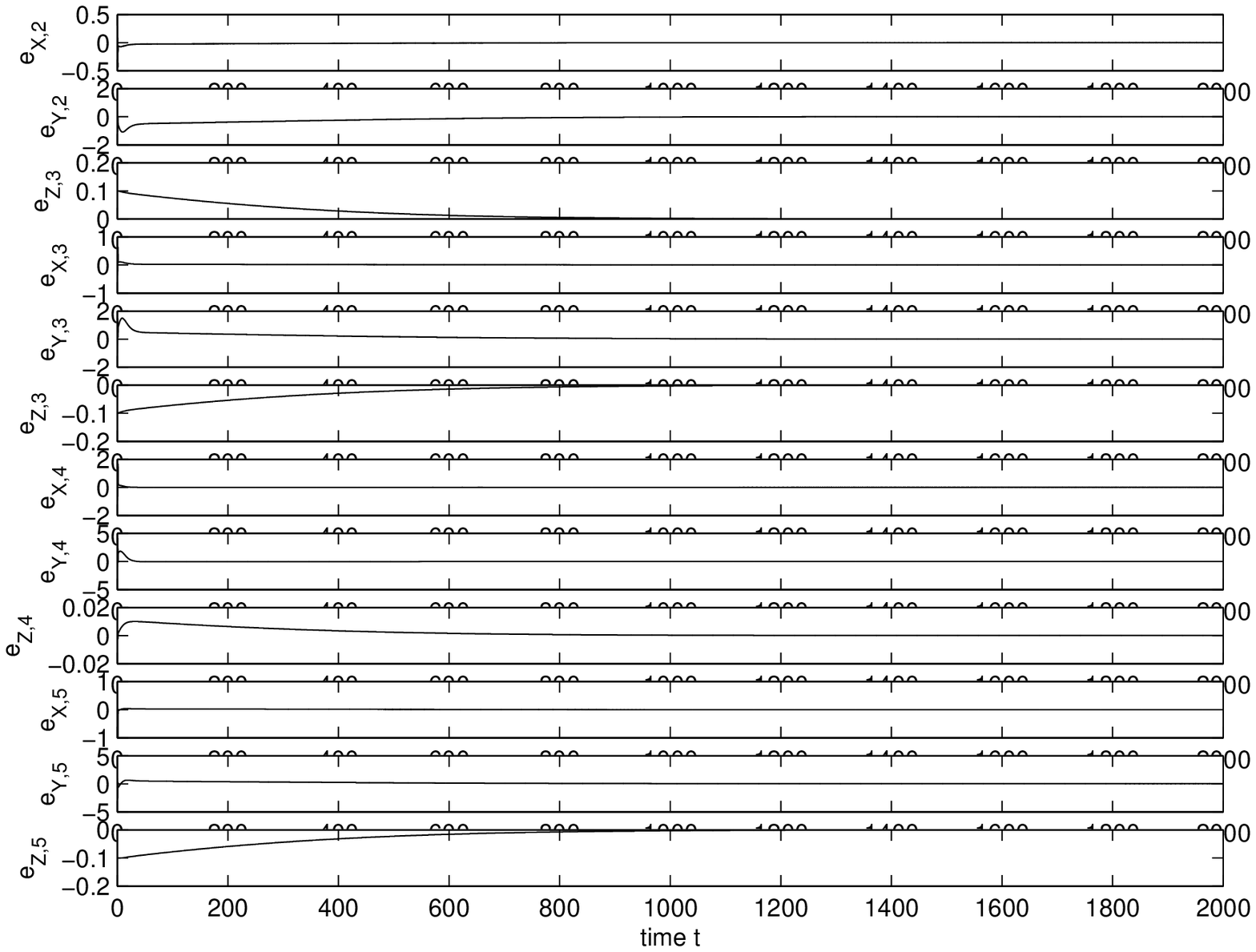,height=6cm,width=6cm}\end{center}
 \begin{center}\begin{minipage}{10cm}{\footnotesize {\bf
 Figure.1}. The collective synchronization emerging in the cycle-type neural network
 (6).
 The collective synchronization in the simply model with 5-coupled chaotic neurons is shown by calculating
synchronization error in system (6), where $e_{X,i}\equiv
X_i-X_1$, $e_{Y,i}\equiv Y_i-Y_1$ and $e_{Z,i}\equiv Z_i-Z_1,
i=2,3,4,5$ respectively denoting the synchronization errors in
each variable converge to zero. Here the initial values of
variables are set as
$(0.2,3,0.7,0.1,4,0.8,0.3,2,0.6,0.1,2,0.7,0.3,4,0.6).$
}\end{minipage}\end{center}
\end{figure}

\begin{figure}
\begin{center}\epsfig{file=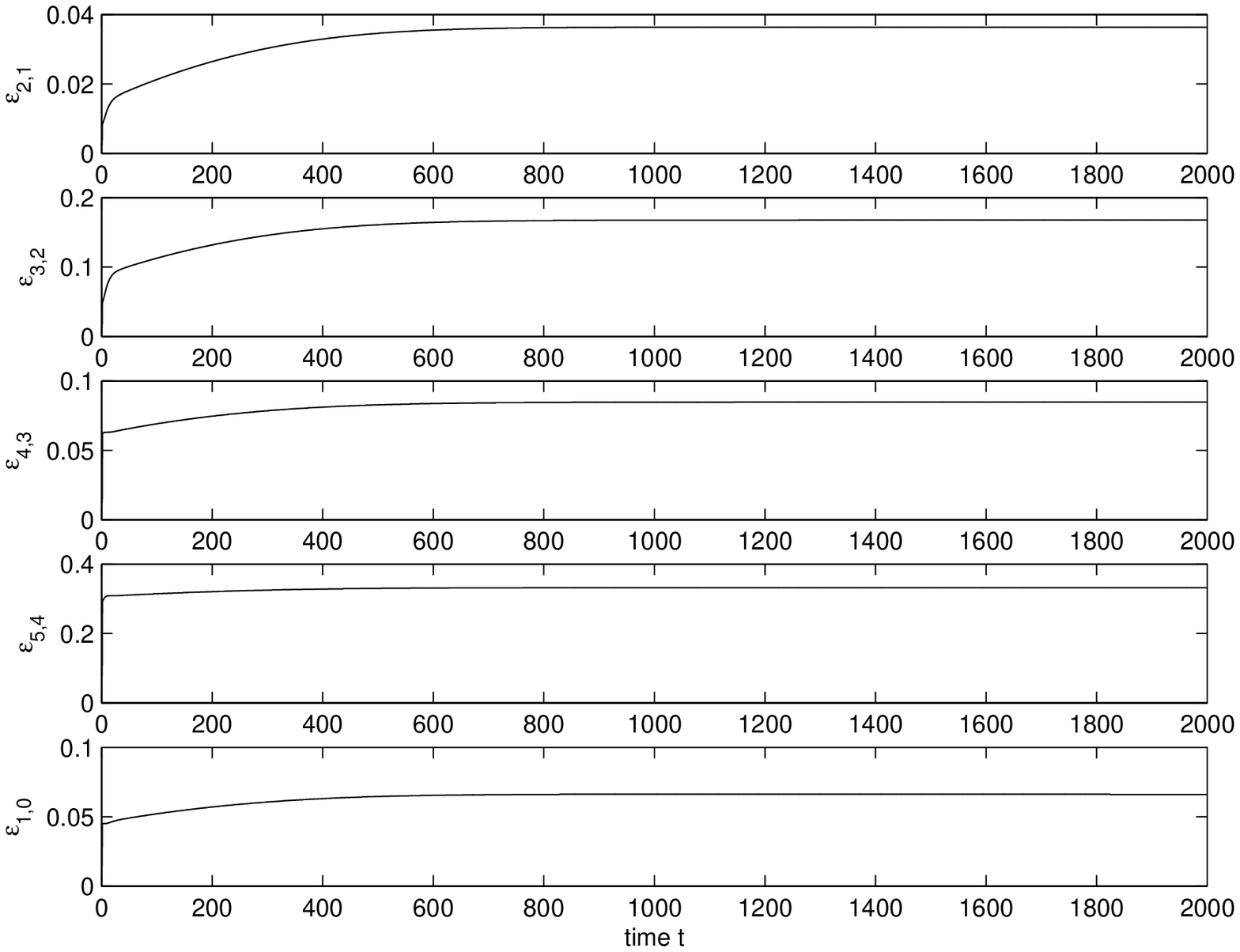,height=6cm,width=6cm}\end{center}
 \begin{center}\begin{minipage}{10cm}{\footnotesize {\bf Figure 2}.
The temporal evolution of the mutual synaptic coupling strength
$\epsilon_{i,i-1},i=1,2,3,4,5$ in system (6)-(7). The mutual
synaptic coupling strength between neurons tends eventually to a
small constant in the course of arising the collective
synchronization in Fig. 1. The evolution is governed by
 the adaptive law (7), where the dissipation parameter is uniformly chosen as
  $\gamma_{i,i-1}=0.1, i=1,2,3,4,5$, and the initial coupling strength is uniformly set as $0$.}\end{minipage}\end{center}
\end{figure}

\begin{figure}
\begin{center}\epsfig{file=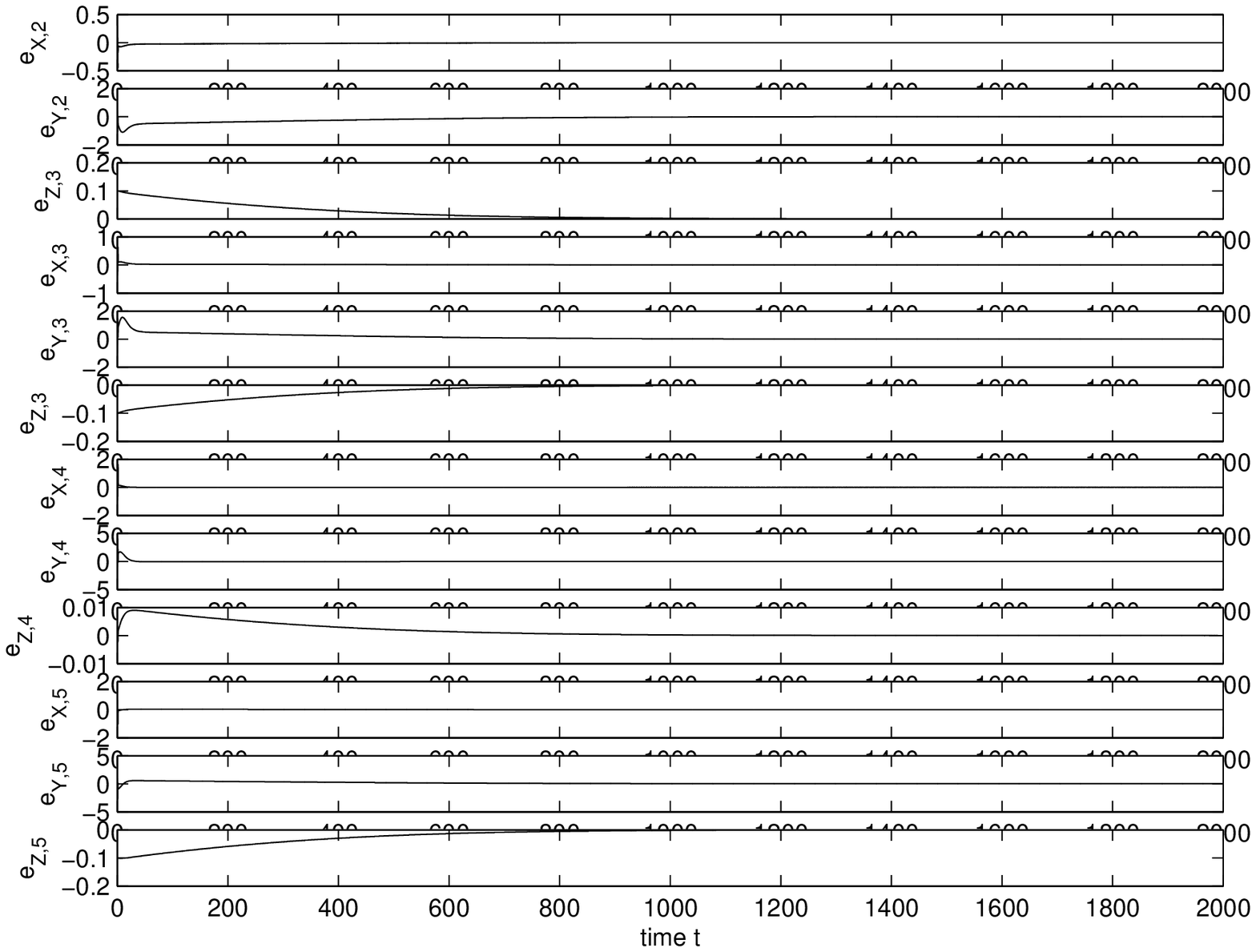,height=6cm,width=6cm}\end{center}
 \begin{center}\begin{minipage}{10cm}{\footnotesize {\bf Figure
 3}. The collective synchronization emerging in the star-type network
 (8) with 5-coupled neurons.
 The collective synchronization in the star-type network of coupled Hindmarsh-Rose chaotic neurons is shown
  by numerically calculating
 synchronization error in system (8), where $e_{X,i}\equiv X_i-X_1$, $e_{Y,i}\equiv
Y_i-Y_1$ and $e_{Z,i}\equiv Z_i-Z_1, i=2,3,4,5$ respectively
denoting the synchronization errors in each variable tend to zero.
Here the initial values of variables are set as
$(0.2,3,0.7,0.1,4,0.8,0.3,2,0.6,0.1,2,0.7,0.3,4,0.6)$.}\end{minipage}\end{center}
\end{figure}

\begin{figure}
\begin{center}\epsfig{file=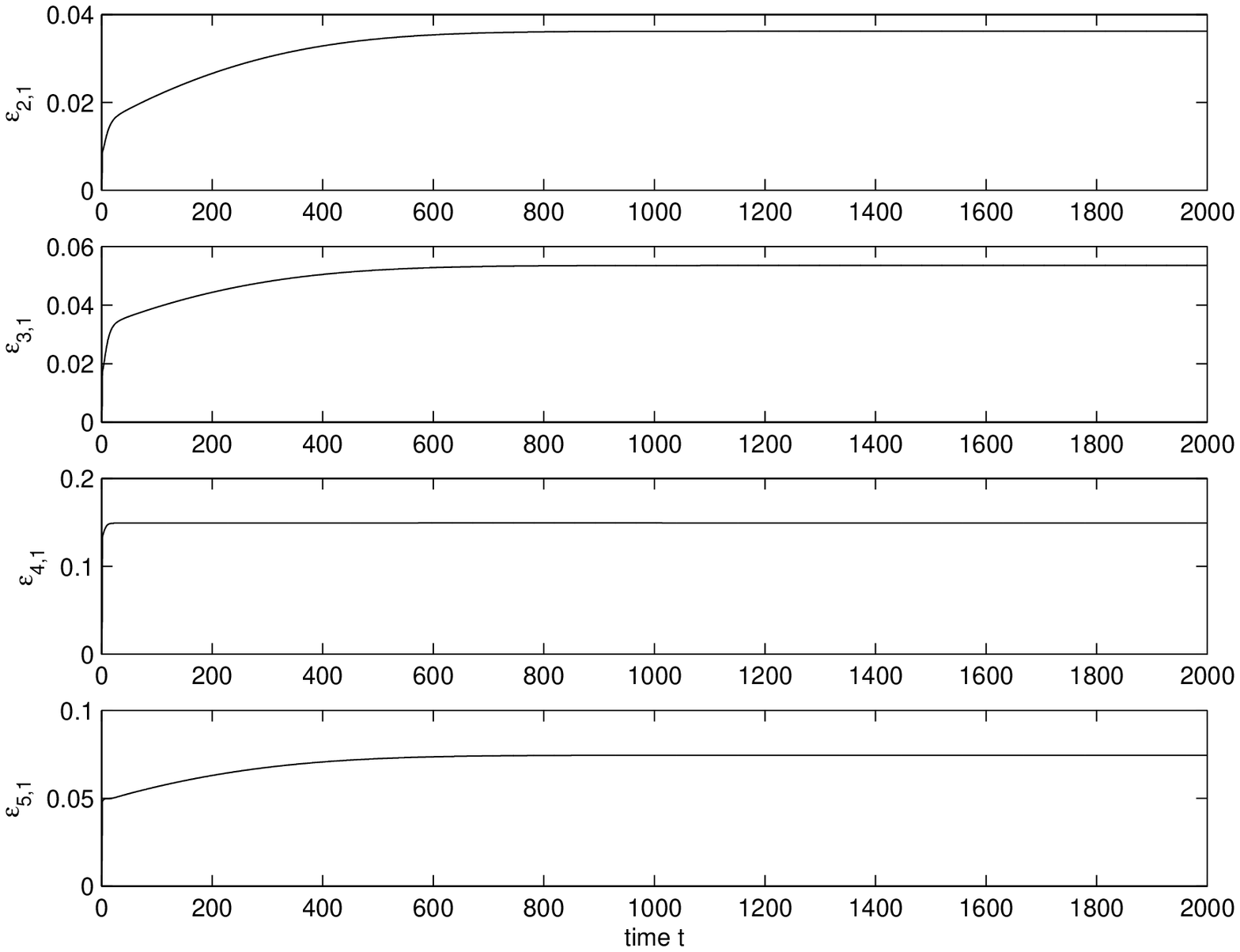,height=6cm,width=6cm}\end{center}
 \begin{center}\begin{minipage}{10cm}{\footnotesize {\bf Figure 4}.
The temporal evolution of the mutual synaptic coupling strength
$\epsilon_{i,1},i=2,3,4,5$ in system (8)-(9). According to the
update law (9) the mutual synaptic coupling intensity between
 individual neurons and the hub neuron self-adaptively approaches
the threshold for producing the collective synchronization in
 Fig. 3,
  where the dissipation parameter is uniformly chosen as
  $\gamma_{i,1}=0.1, i=2,3,4,5$, and the initial coupling strength is uniformly set as
  $0$. Obviously here mean coupling strength is smaller than
  that in Fig. 2.
}\end{minipage}\end{center}
\end{figure}


{\small
 \begin{thebibliography}{14.8cm}
\bibitem {s93} S.H. Strogatz and I. Stewart, {\sl Sic. Am.} {\bf 269}, 102 (1993); D. Helbing, and
B.A. Huberman, {\sl Nature} {\bf 396}, 738 (1998); A.T. Winfree,
{\sl The Geometry of Biology Time} (2nd edn, Sporinger, New York,
2001); I.Z. Kiss, Y. Zhai, and J.L. Hudson, {\sl Science} {\bf
296}, 1676 (2002); A.T. Winfree, {\sl Science} {\bf 298}, 2336
(2002); S. Nadis, {\sl Nature} {\bf
 421}, 780 (2003); S.H. Strogatz, {\sl Sync: the emerging science of spontaneous order} (Hyperion, New York, 2003).
\bibitem {w61} N. Wiener, {\sl Cybernetics} (2nd edn, MIT Press, Cambridge, Massachusetts, 1961).
\bibitem {w67}A.T. Winfree, {\sl J. Theor. Biol.} {\bf 16}, 15 (1967).
\bibitem {k84} Y. Kuramoto, {\sl Chemical Oscillations, Waves, and Turbulence} (Springer, Berlin, 1984).
\bibitem {h91}E.D. Lumer, and B.A. Huberman, {\sl Phys. Lett. A} {\bf 160}, 227 (1991); S.H. Strogatz, R.E. Mirollo and P.C. Matthews, {\sl Phys. Rev. Lett.} {\bf 68},
2730 (1992); H. Daido, {\sl Phys. Rev. Lett.} {\bf 73}, 760
(1994); J.D. Crawford,  {\sl Phys. Rev. Lett.} {\bf 74}, 4341
(1995); K. Wiesenfeld, P. Colet and S.H. Strogatz,  {\sl Phys.
Rev. Lett.} {\bf 76}, 404 (1996); H. Daido, {\sl Physica D} {\bf
91}, 24 (1996); J.D. Crawford and K.T.R. Davies, {\sl Physica D}
{\bf 125}, 1 (1999); S.H. Strogatz, {\sl Physica D} {\bf 143}, 1
(2000).
\bibitem {h04}Debin Huang, {\sl Phys. Rev. Lett.} {\bf 93}, 214101
(2004).
\bibitem {g89} C.M. Gray,  P. Koning, A.K. Engel and  W. Singer,
 {\sl
Nature} {\bf 338}, 334 (1989); R.C. Elson, and A.I. Selverston,
{\sl J. Neurophysiol.} {\bf 74}, 1996 (1995);  C.M. Gray and  D.A.
McCormick, {\sl Scienc} {\bf 274}, 109 (1996); R.C. Elson, {\sl et
al},  {\sl Phys. Rev. Lett.} {\bf 81}, 5692 (1998);  E.M.
Izhikevich, {\sl SIAM Rev.} {\bf 43}, 315 (2001); J.M. Samonds,
J.D. Allison, H.A. Brown and A.B. Bonds,
 {\sl Proc. Natl. Acad. Sci. USA} {\bf 101},
6722 (2004).
\bibitem {h92}  D. Hansel and H. Sompolinsky, {\sl Phys. Rev. Lett.} {\bf 68}, 718 (1992);
M. Bazhenov, R. Huerta, M.I. Rabinovich and T. Sejnowski, {\sl
Physica D} {\bf 116}, 392 (1998); N.F. Rulkov, {\sl Phys. Rev.
Lett.} {\bf 86}, 183 (2001);  M.G. Rosenblum and A.S. Pikovsky,
{\sl Phys. Rev. Lett.} {\bf 92}, 114102 (2004).
\bibitem {h84}J.L. Hindmarsh and R.M. Rose,
{\sl Proc. R. Soc. London B} {\bf 221}, 87 (1984).
\bibitem {s98} D.J. Watts and S.H. Strogatz, {\sl Nature} {\bf 393}, 440
(1998); S.H. Strogatz, {\sl Nature} {\bf 410}, 268 (2001).
\bibitem {h49}D. Hebb, {\sl The organization of behavior} (New York: Wiley, 1949).
\bibitem {b99}  A.L. Barabasi and L. Albert, {\sl Science} {\bf 286}, 509 (1999);
 L. Albert and A.L. Barabasi, {\sl Rev. Mod. Phys.} {\bf 74},
47 (2002).
\end{thebibliography}}
\end{document}